\documentclass[prl,aps,twocolumn,superscriptaddress,showpacs]{revtex4}
\usepackage{graphicx}

\begin{document}

\title{       Evolution of Spin-Orbital-Lattice Coupling
              in the $R$VO$_3$ Perovskites }

\author {     Peter Horsch }
\affiliation{ Max-Planck-Institut f\"ur Festk\"orperforschung,
              Heisenbergstrasse 1, D-70569 Stuttgart, Germany }

\author {     Andrzej M. Ole\'{s} }
\affiliation{ Max-Planck-Institut f\"ur Festk\"orperforschung,
              Heisenbergstrasse 1, D-70569 Stuttgart, Germany }
\affiliation{ Marian Smoluchowski Institute of Physics, Jagellonian
              University, Reymonta 4, PL-30059 Krak\'ow, Poland }

\author {     Louis Felix Feiner }
\affiliation{ Philips Research Laboratories, High Tech Campus 4,
              NL-5656 AE Eindhoven, The Netherlands }

\author {     Giniyat Khaliullin }
\affiliation{ Max-Planck-Institut f\"ur Festk\"orperforschung,
              Heisenbergstrasse 1, D-70569 Stuttgart, Germany }

\date{4 December 2007; published 23 April 2008}

\begin{abstract}
We introduce a microscopic model which unravels the physical
mechanisms responsible for the observed phase diagram of the
$R$VO$_3$ perovskites. It reveals a nontrivial interplay between
superexchange, the orbital-lattice coupling due to the
GdFeO$_3$-like rotations of the VO$_6$ octahedra, and orthorhombic
lattice distortions. We find that the lattice strain affects the
onset of the magnetic and orbital order by partial suppression of
orbital fluctuations. The present approach provides also a natural
explanation of the observed reduction of magnon energies from
LaVO$_3$ to YVO$_3$.\\
{\it Published in: Phys. Rev. Lett. {\bf 100}, 167205 (2008).}
\end{abstract}


\maketitle

%%%%%%%%%%%%%%%%%%%%%%%%%%%%%%%%%%%%%%%%%%%%%%%%%%%%%%%%%%%%
%%                coupling between OO and SO
%%%%%%%%%%%%%%%%%%%%%%%%%%%%%%%%%%%%%%%%%%%%%%%%%%%%%%%%%%%%
Over the last decade, extensive work on transition metal oxides
has demonstrated a strong interrelationship between spin order (SO)
and orbital order (OO), often compounded by the occurrence of various
types of lattice distortions, resulting in phase behavior of dazzling
complexity. Recently, however, impressive experimental work has
produced exceptionally detailed information on the phase diagrams of
the $R$MnO$_3$ manganites \cite{Goo06} and the $R$VO$_3$ vanadates
(where $R$=Lu,Yb,$\cdots$,La) \cite{Miy03}, thus providing a unique
challenge to the theory and the opportunity to resolve the interplay
between the underlying microscopic mechanisms.

%%%%%%%%%%%%%%%%%%%%%%%%%%%%%%%%%%%%%%%%%%%%%%%%%%%%%%%%%%%%
%%               manganites versus vanadates
%%%%%%%%%%%%%%%%%%%%%%%%%%%%%%%%%%%%%%%%%%%%%%%%%%%%%%%%%%%%
The manganite $R$MnO$_3$ perovskites exhibit the more common behavior,
i.e., upon lowering the temperature, the OO appears first, accompanied
by a lattice distortion, at $T_{\rm OO}$, and thus {\it modifies\/}
the conditions for the onset of SO at a significantly lower temperature
$T_{\rm N}$. When the ionic radius $r_R$ of the $R^{3+}$ ions decreases,
the N\'eel temperature $T_N$ gets drastically reduced and the OO
transition temperature $T_{\rm OO}$ is enhanced
\cite{Goo06}. By contrast, in the $R$VO$_3$ vanadates the two
transitions are close to each other, i.e., $T_{N1}\lesssim T_{\rm OO}$,
the type of order being $G$-type OO ($G$-OO) and $C$-type
antiferromagnetic ($C$-AF), setting in below $T_{\rm OO}$ and $T_{N1}$
\cite{Miy06}, respectively \cite{notecg}. Moreover, whereas $T_{N1}$
again gets reduced for decreasing $r_R$, $T_{\rm OO}$ exhibits a
{\it nonmonotonic\/} dependence on $r_R$ \cite{Miy03}.

%%%%%%%%%%%%%%%%%%%%%%%%%%%%%%%%%%%%%%%%%%%%%%%%%%%%%%%%%%%%
%%              complexity of vanadates
%%%%%%%%%%%%%%%%%%%%%%%%%%%%%%%%%%%%%%%%%%%%%%%%%%%%%%%%%%%%
These experimental results demonstrate that spins and orbitals are
intimately coupled in the $R$VO$_3$ vanadates, consistent with the
recent observation that these compounds form a unique class
characterized by {\it strong orbital fluctuations\/}
\cite{Ulr03,Miy05,And07} which follow from superexchange interactions
between almost degenerate $t_{2g}$ orbitals \cite{Kha01,Ole07} and
spin-orbit term \cite{Hor03,Goo07}. Their coupling is also
visible in spectacular changes of the SO and OO under pressure
\cite{Goo07}. The pressure dependence and thermal conductivity data
\cite{Yan07} suggest in turn {\it strong orbital-lattice coupling\/}
\cite{Goo04}. As in $t_{2g}$ systems Jahn-Teller (JT) interactions
are expected to be rather weak, the GdFeO$_3$-like distortions (GFOD)
\cite{Ima04,Pav05} are the prime candidate for being involved in the
coupling between orbitals and the lattice.

\begin{figure}[b!]
\includegraphics[width=7.2cm]{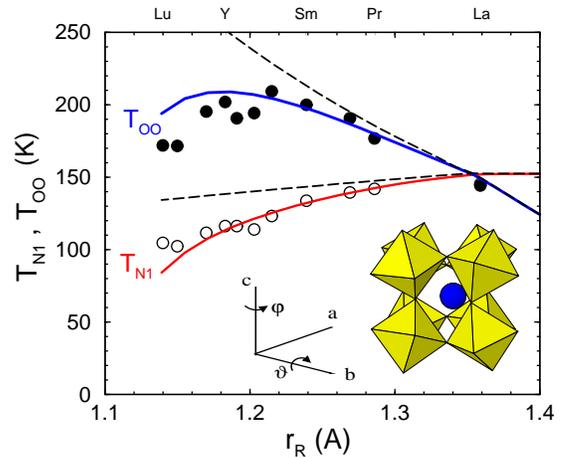}
\caption{(color)
The orbital transition $T_{\rm OO}$ and N\'eel $T_{N1}$ temperature
(solid lines) for varying ionic size in $R$VO$_3$, as obtained from
the present theory for the parameter values given in the text, and
from experiment (full and empty circles) \cite{Miy03}. Dashed lines
indicate $T_{\rm OO}$ and $T_{N1}$ obtained under neglect of
orbital-lattice coupling ($g_{\rm eff}=0$). The inset shows the
GdFeO$_3$-type distortion, with the rotation angles $\vartheta$ and
$\varphi$ corresponding to the data of YVO$_3$ \cite{Ree06}.
}
\label{fig:phd}
\end{figure}

%%%%%%%%%%%%%%%%%%%%%%%%%%%%%%%%%%%%%%%%%%%%%%%%%%%%%%%%%%%%
%%                    this Letter
%%%%%%%%%%%%%%%%%%%%%%%%%%%%%%%%%%%%%%%%%%%%%%%%%%%%%%%%%%%%
In this Letter we present a first microscopic approach to the phase
diagram of the $R$VO$_3$ vanadates using an extended spin-orbital model
which treats the coupled spin and orbital degrees of freedom in the
presence of lattice distortions. We demonstrate that the generic trends
observed in the phase diagram of $R$VO$_3$ can be indeed explained by
the theory, see Fig. \ref{fig:phd},
provided one includes explicitly the coupling between the orbitals and
the lattice distortions that increase with decreasing $r_R$.

%%%%%%%%%%%%%%%%%%%%%%%%%%%%%%%%%%%%%%%%%%%%%%%%%%%%%%%%%%%%
%%                      fig. 1
%%%%%%%%%%%%%%%%%%%%%%%%%%%%%%%%%%%%%%%%%%%%%%%%%%%%%%%%%%%%

%%%%%%%%%%%%%%%%%%%%%%%%%%%%%%%%%%%%%%%%%%%%%%%%%%%%%%%%%%%%
%%                      role of tilting
%%%%%%%%%%%%%%%%%%%%%%%%%%%%%%%%%%%%%%%%%%%%%%%%%%%%%%%%%%%%
{\it A priori\/}, the decreasing ionic radius $r_R$ in the $R$VO$_3$
perovskites triggers the GFOD obtained by alternating rotations of
the VO$_6$ octahedra by an angle $\vartheta$ around the $b$ axis,
and by an angle $\varphi$ around the $c$ axis
(see inset in Fig. \ref{fig:phd}). This results in a decrease of
V--O--V bond angles (e.g. $\Theta=\pi-2\vartheta$ along the $c$
direction), and leads to an orthorhombic distortion $u=(b-a)/a$,
where $a$ and $b$ are the lattice parameters of the $Pbnm$ structure.
Although these changes are systematic in the oxides \cite{Goo05},
their relation to the evolution of microscopic parameters and physical
properties such as the onset of OO and SO along the $R$VO$_3$ series
is not yet fully understood.

%%%%%%%%%%%%%%%%%%%%%%%%%%%%%%%%%%%%%%%%%%%%%%%%%%%%%%%%%%%%
%%                   full vanadate model
%%%%%%%%%%%%%%%%%%%%%%%%%%%%%%%%%%%%%%%%%%%%%%%%%%%%%%%%%%%%
The spin-orbital model for $R$VO$_3$ includes: (i) the
superexchange interaction \cite{Kha01}, (ii) the crystal field
(CF) splitting which follows from the GFOD, (iii) orbital-orbital
interactions induced by orbital-lattice coupling, (iv)
orbital-strain coupling. It takes the generic form
\begin{eqnarray}
\label{som}
{\cal H}&=&\!J\sum_{\langle ij\rangle}\!
\left\{\!\Big({\vec S}_i\!\cdot\!{\vec S}_j\!+\!S^2\Big){{\cal
J}}_{ij} + {{\cal K}}_{ij}\!\right\} -
V_{c}(\vartheta)\!\sum_{\langle ij\rangle\parallel c}\!
\tau_i^z\tau_j^z
\nonumber \\
&+&E_z(\vartheta)\,\sum_i\!e^{i{\vec R}_i{\vec Q}}\tau_i^z
+ V_{ab}(\vartheta)\!\sum_{\langle ij\rangle\perp\; c}\tau_i^z\tau_j^z
+ {\cal H}_u,
\end{eqnarray}
where the parameters $\{E_z,V_{ab},V_c\}$ depend on the tilting
angle $\vartheta$. The first term describes the superexchange of
strength $J=4t^2/U$ [here $t$ is the effective $(dd\pi)$ hopping
between $t_{2g}$ orbitals and $U$ the intraorbital Coulomb
interaction] between $V^{3+}$ ions in the $d^2$ configuration with
$S=1$ spins. The dependence of $J$ on the rare earth ion $R$ is
weak \cite{Goo06}, and is neglected in the present theory; we
adopted $J=202$ K for the theoretical curves in Fig.
\ref{fig:phd}. The orbital operators ${\cal J}_{ij}$ and ${\cal
K}_{ij}$ follow from virtual $d^2_id^2_j\rightarrow d^3_id^1_j$
charge excitations and depend on Hund's exchange parameter
$J_H/U$. Their form depends on the $\langle ij\rangle$-bond
orientation; they are given in Ref. \cite{Ole07} for the actual
$(xy)^1(yz/zx)^1$ configuration in cubic vanadates. The orbital
(pseudospin) operators $\tau_i^z\equiv
\frac{1}{2}(n_{yz}-n_{zx})_i$ refer to the two active orbitals:
$yz$ and $zx$ \cite{Kha01,Ole07}. The CF splitting of these two
orbitals $\propto E_z$ supports $C$-type OO \cite{notecg}, with a
modulation vector ${\vec Q}=(\pi,\pi,0)$ in cubic notation. The
$V_{ab}>0$ and $V_c>0$ orbital interactions are due to the JT and
GFOD distortions of the VO$_6$ octahedra, and like $E_z$ favor
$C$-type OO. Unlike for $V_{ab}$, it may be expected that the
dependence of $V_{c}$ on the angle $\vartheta$ is weak, and indeed
a constant $V_{c}(\vartheta)=0.26J$ reproduces a simultaneous
onset of SO and OO in LaVO$_3$ within the present model
\cite{notevc}, i.e., $T_{\rm OO}=T_{N1}$, see Fig. \ref{fig:phd}.
Finally, ${\cal H}_u$ describes the coupling of the orbitals to
the orthorhombic distortion $u$ and is explained below.

%%%%%%%%%%%%%%%%%%%%%%%%%%%%%%%%%%%%%%%%%%%%%%%%%%%%%%%%%%%%
%%                   dependence on theta
%%%%%%%%%%%%%%%%%%%%%%%%%%%%%%%%%%%%%%%%%%%%%%%%%%%%%%%%%%%%
To derive the functional dependence of the microscopic parameters
$\{E_z,V_{ab}\}$ on the tilting angle $\vartheta$, we considered
the point charge model, and used the structural data for $R$VO$_3$
\cite{notece}. Due to the GFOD shown in Fig. \ref{fig:phd}, the
two active $yz/zx$ orbitals experience the CF splitting $E_z$. By
considering the ionic charges acting on the $t_{2g}$ orbitals and
analyzing the level splittings, we obtained:
\begin{equation}
\label{Ez}
E_z(\vartheta)=J\,v_z\,\sin^3\vartheta\cos\vartheta,
\end{equation}
and verified that the $xy$ orbitals are indeed well below the
$\{yz,zx\}$ orbitals. These qualitative trends are valid in a
range of $\varphi$, and for further analysis we adopted a
representative value $\varphi=\vartheta/2$, similar to the trend
in titanates \cite{Pav05}. It is expected that the angular
dependence of the orbital interaction $V_{ab}$ follows the CF term
(\ref{Ez}):
\begin{equation}
\label{vab}
V_{ab}(\vartheta)=J\,v_{ab}\,\sin^3\vartheta\cos\vartheta.
\end{equation}

%%%%%%%%%%%%%%%%%%%%%%%%%%%%%%%%%%%%%%%%%%%%%%%%%%%%%%%%%%%%
%%                  transverse orbital field
%%%%%%%%%%%%%%%%%%%%%%%%%%%%%%%%%%%%%%%%%%%%%%%%%%%%%%%%%%%%
An important term in (\ref{som}), coupling the orbitals to the
lattice, is the one involving the orthorhombic strain $u$,
\begin{equation}
\label{Hol}
{\cal H}_u\equiv -\;gu\sum_i\tau_i^x+\frac12 N K(u-u_0(\vartheta))^2,
\end{equation}
where $g>0$ is the coupling constant, $K$ is the force constant,
and $N$ is the number of $V^{3+}$ ions. In contrast to the
longitudinal field $E_z$, $gu$ acts as a transverse field, i.e.,
it favors that one of the two linear combinations
$\frac{1}{\sqrt{2}}(|yz\rangle\pm|zx\rangle)$ is occupied. Since
$u$ is a classical variable, we may minimize Eq. (\ref{Hol}) and
write the global distortion  as $u(\vartheta;T)\equiv
u_0(\vartheta)+(g/K)\langle\tau^x\rangle_T$, consisting of (i) a
pure lattice contribution $u_0(\vartheta)$, and (ii) a
contribution due the orbital polarization
$\propto\langle\tau^x\rangle$ which we determined
self-consistently. Both $u_0$ and $\langle\tau^x\rangle$ are
expected to increase with increasing tilting $\vartheta$. As will
be shown below, $\langle\tau^x\rangle$ has only a weak temperature
dependence, so we may use
\begin{equation}
\label{geff}
g_{\rm eff}(\vartheta)\equiv gu(\vartheta)=
J\,v_{g}\,\sin^5\vartheta\cos\vartheta.
\end{equation}
Indeed, we shall see below that this strong dependence of $g_{\rm
eff}$ on $\vartheta$ is not only necessary to reduce $T_{\rm OO}$
for large tilting angles $\vartheta$, but is also consistent with
the experimental data for $u(\vartheta)$
\cite{Ren03,Ree06,Sag06,Sag07}. Altogether, $\{v_z,v_{ab},v_g\}$
are the parameters of the spin-orbital model (\ref{som}).

%%%%%%%%%%%%%%%%%%%%%%%%%%%%%%%%%%%%%%%%%%%%%%%%%%%%%%%%%%%%
%%                     phase diagram
%%%%%%%%%%%%%%%%%%%%%%%%%%%%%%%%%%%%%%%%%%%%%%%%%%%%%%%%%%%%
In the $R$MnO$_3$ manganites, mean-field (MF) theory with
classical, on-site, SO and OO parameters can be used to
investigate the phase diagram \cite{Fei99}. However, this approach
with on-site order parameters does not suffice in the vanadates
\cite{Silva} when orbital fluctuations stabilizing the $C$-AF
phase are present \cite{Kha01} --- then it becomes essential to
determine self-consistently the orbital singlet correlations
$\langle{\vec\tau}_i\cdot{\vec\tau}_j\rangle$ as well. Hence we
used a cluster MF theory for a bond $\langle ij\rangle$ along the
$c$ axis \cite{noteos}, with spin and orbital MF terms $\langle
S^z\rangle$ and
$\langle\tau^z\rangle_G\equiv\frac12|\langle\tau^z_i-\tau^z_j\rangle|$
which follow from interactions with the V$^{3+}$ neighbors in all
three cubic directions. Apart from the singlet orbital
correlations $\langle{\vec\tau}_i\cdot{\vec\tau}_j\rangle$,  the
transverse field $\langle\tau^x\rangle$ was crucial to reproduce
the phase diagram of Fig. \ref{fig:phd}.

%%%%%%%%%%%%%%%%%%%%%%%%%%%%%%%%%%%%%%%%%%%%%%%%%%%%%%%%%%%%
%%                      fig. 2
%%%%%%%%%%%%%%%%%%%%%%%%%%%%%%%%%%%%%%%%%%%%%%%%%%%%%%%%%%%%
\begin{figure}[t!]
\includegraphics[width=6.2cm]{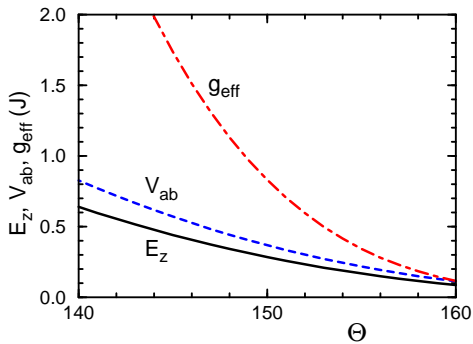}
\caption{(color online)
Parameters $\{E_z,V_{ab},g_{\rm eff}\}$ in Eq. (\ref{som}) (in
units of $J$) for varying V-O-V bond angle $\Theta$ (in degrees).
Parameters: $v_z=17$, $v_{ab}=22$, $v_{g}=740$.
}
\label{fig:lat}
\end{figure}

%%%%%%%%%%%%%%%%%%%%%%%%%%%%%%%%%%%%%%%%%%%%%%%%%%%%%%%%%%%%
%%                  variation of parameters
%%%%%%%%%%%%%%%%%%%%%%%%%%%%%%%%%%%%%%%%%%%%%%%%%%%%%%%%%%%%
The nontrivial character of the phase diagram and the underlying
spin-orbital coupling in the $R$VO$_3$ vanadates can be fully
appreciated by analyzing the variation of the microscopic interactions
with decreasing angle $\Theta$ (increasing tilting $\vartheta$). While
the CF splitting and the orbital interactions $V_{ab}$ show only
moderate increase for decreasing $\Theta$, the orbital polarization
$g_{\rm eff}$ increases rapidly and becomes quite large when
$\Theta<150^{\circ}$ (Fig. \ref{fig:lat}). Note that the increasing JT
term $V_{ab}$ supports the superexchange and stabilizes the $G$-OO,
both the increasing CF splitting $E_z$, and the orbital-lattice
coupling $g_{\rm eff}$ compete with it.

\begin{figure}[t!]
\includegraphics[width=6.4cm]{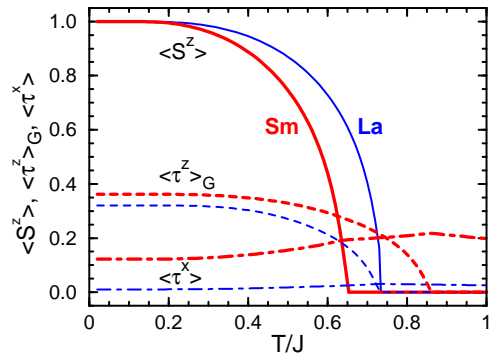}
\caption{(color online)
Spin $\langle S^z\rangle$ (solid) and $G$-type orbital
$\langle\tau^z\rangle_G$ (dashed) order parameters, vanishing at
$T_{\rm OO}$ and $T_{N1}$, respectively, and the transverse orbital
polarization $\langle\tau^x\rangle$ (dashed-dotted lines)
for LaVO$_3$ and SmVO$_3$ (thin and heavy lines) for $V_{c}=0.26J$;
other parameters as in Fig. \ref{fig:lat}.
}
\label{fig:ops}
\end{figure}

%%%%%%%%%%%%%%%%%%%%%%%%%%%%%%%%%%%%%%%%%%%%%%%%%%%%%%%%%%%%
%%                      OO and SO
%%%%%%%%%%%%%%%%%%%%%%%%%%%%%%%%%%%%%%%%%%%%%%%%%%%%%%%%%%%%
With the present parameters OO and SO occur simultaneously in
LaVO$_3$, and $T_{\rm OO}=T_{N1}\simeq 0.73J$ (Fig.
\ref{fig:ops}). The orbital polarization
$\langle\tau^x\rangle\simeq 0.03$ is here rather weak at $T_{N1}$,
and is further reduced in the ordered phase when the OO parameter
$\langle\tau^z\rangle_G$ grows with decreasing $T<T_{\rm OO}$ (due
to finite $E_z$, the orbitals $xz/zy$ are nonequivalent and
$\langle\tau^z\rangle_0\equiv |\langle\tau^z_i\rangle|>0$ even for
$T>T_{\rm OO}$ \cite{notetz}). In contrast, in SmVO$_3$ the OO
occurs first at $T_{\rm OO}\simeq 0.86J$, and the SO follows only
at $T_{N1}\simeq 0.65J$. For these parameters the transverse
orbital polarization is considerably increased, with
$\langle\tau^x\rangle\simeq 0.20$ at $T_{N1}$ (see Fig.
\ref{fig:ops}). Note that the polarization $\langle\tau^x\rangle$
does not change at $T\simeq T_{\rm OO}$, and only below $T_{N1}$
there is a weak reduction of $\langle\tau^x\rangle$, in agreement
with experiment \cite{Sag07}. In both cases the $G$-OO parameter
$\langle\tau^z\rangle_G$ is reduced by singlet orbital
fluctuations, being $\langle\tau^z\rangle_G\simeq 0.32$ (0.37) for
LaVO$_3$ (SmVO$_3$).

\begin{figure}[b!]
\includegraphics[width=6.0cm]{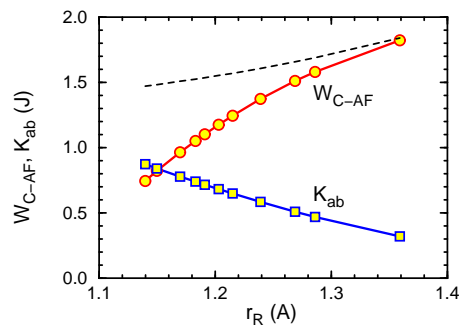}
\caption{(color online)
The width of magnon band $W_{C-{\rm AF}}$ for finite $g_{\rm eff}$
(circles) and without orbital-strain coupling ($g_{\rm eff}=0$,
dashed), and orbital interactions in $ab$ planes $K_{ab}$ (squares)
in the $C$-AF phase of cubic vanadates (the points correspond to
the $R$VO$_3$ compounds of Fig. \ref{fig:phd}).
Parameters as in Fig. \ref{fig:lat}.
}
\label{fig:all}
\end{figure}

As a result of the competition between the JT term and the CF and
orbital-lattice interaction, the temperature $T_{\rm OO}$ {\it
increases\/} first only moderately with decreasing $r_R$ and next
{\it decreases\/}, resulting in two distinct regimes of the phase
diagram of Fig. \ref{fig:phd}. First, when $\Theta$ decreases from
$157.4^{\circ}$ in LaVO$_3$ to $144.8^{\circ}$ in YVO$_3$,
increasing $V_{ab}$ dominates and $T_{\rm OO}$ increases (Fig.
\ref{fig:phd}). This is similar to the $R$MnO$_3$ manganites
\cite{Goo06} and can be understood by considering the {\it
total\/} orbital interactions $K_{ab}\tau^z_i\tau^z_j$ in the $ab$
planes, including both the superexchange $J$ and the JT term
$V_{ab}$, see Fig. \ref{fig:all}. Here we use again the ionic
radius $r_R$ as in Fig. \ref{fig:phd} --- we deduced its relation
to the tilting angle $\vartheta$,
$r_R=r_0-\alpha\sin^22\vartheta)$ with $r_0=1.5$ \AA{} and
$\alpha=0.95$ \AA{}, from the structural data of Refs.
\cite{Ren03,Ree06,Sag06,Sag07}. Note that the CF splitting $E_z$
increases with decreasing $r_R$, so it partly compensates the
effect of increasing $V_{ab}$. Second, the rapidly increasing
orbital polarization $g_{\rm eff}$ (Fig. \ref{fig:lat}) suppresses
the tendency towards $G$-OO and triggers the observed drop of
$T_{\rm OO}$ (Fig. \ref{fig:phd}) when $r_R$ decreases beyond
$r_R\sim 1.18$ \AA{} found in YVO$_3$.

%%%%%%%%%%%%%%%%%%%%%%%%%%%%%%%%%%%%%%%%%%%%%%%%%%%%%%%%%%%%
%%                      fig. 3
%%%%%%%%%%%%%%%%%%%%%%%%%%%%%%%%%%%%%%%%%%%%%%%%%%%%%%%%%%%%

%%%%%%%%%%%%%%%%%%%%%%%%%%%%%%%%%%%%%%%%%%%%%%%%%%%%%%%%%%%%
%%                      fig. 4
%%%%%%%%%%%%%%%%%%%%%%%%%%%%%%%%%%%%%%%%%%%%%%%%%%%%%%%%%%%%

%%%%%%%%%%%%%%%%%%%%%%%%%%%%%%%%%%%%%%%%%%%%%%%%%%%%%%%%%%%%
%%                 reduction of magnon energy
%%%%%%%%%%%%%%%%%%%%%%%%%%%%%%%%%%%%%%%%%%%%%%%%%%%%%%%%%%%%
The changes in orbital correlations caused by the lattice-induced
increase of the total orbital interactions $K_{ab}$ with
decreasing $r_R$ (see Fig. \ref{fig:all}) suppress the magnetic
interactions in the $C$-AF phase, so the total magnon energy scale
$W_{C-{\rm AF}}=4(J_{ab}+|J_c|)$ (at $T=0$) \cite{Ole07} is
reduced from $\sim 1.84J$ in LaVO$_3$ to $\sim 1.05J$ in YVO$_3$,
i.e., by a factor close to 1.8. This explains qualitatively the
rather small magnon energies observed in the $C$-AF phase of
YVO$_3$ \cite{Ulr03}. The reduction is at first instance
surprising as the value of $J$ does not change at all, and it is
caused solely by the suppression of the singlet orbital
correlations $\langle{\vec\tau}_i\cdot{\vec\tau}_j\rangle$ by the
transverse field $g_{\rm eff}(\vartheta)$ (while this effect is
small for $g_{\rm eff}=0$, in conflict with experiment).

%%%%%%%%%%%%%%%%%%%%%%%%%%%%%%%%%%%%%%%%%%%%%%%%%%%%%%%%%%%%
%%                      fig. 5
%%%%%%%%%%%%%%%%%%%%%%%%%%%%%%%%%%%%%%%%%%%%%%%%%%%%%%%%%%%%
\begin{figure}[t!]
\includegraphics[width=8.2cm]{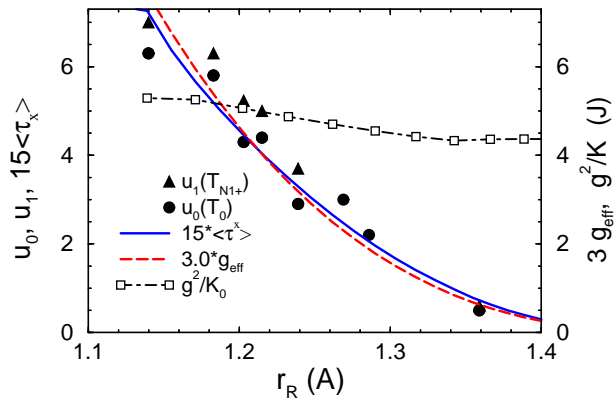}
\caption{(color online)
Experimental distortion (in percent) at $T_0=0$ ($u_0$, circles)
and above $T_{N1}$ ($u_1$, triangles) for LaVO$_3$ \cite{Ren03}
and other $R$VO$_3$ compounds \cite{Ree06,Sag07},
compared with the orbital polarization
$\langle\tau^x\rangle_{T=0}$ and with $g_{\rm eff}$ (\ref{geff});
$g_{\rm eff}$ and $g^2/K$ are in units of $J$.
Squares show the upper bound for $g^2/K$ predicted by the theory
(at $u_0=0$).
Parameters as in Fig. \ref{fig:lat}.
}
\label{fig:pol}
\end{figure}

%%%%%%%%%%%%%%%%%%%%%%%%%%%%%%%%%%%%%%%%%%%%%%%%%%%%%%%%%%%%
%%        unexpected outcome on lattice distortions
%%%%%%%%%%%%%%%%%%%%%%%%%%%%%%%%%%%%%%%%%%%%%%%%%%%%%%%%%%%%
The role played by the orbital-strain coupling (\ref{Hol}) in the
phase diagram of the $R$VO$_3$ compounds becomes even more transparent
by comparing the dependence of $g_{\rm eff}$ on the ionic radius $r_R$
with the actual lattice distortion $u$ in $R$VO$_3$ (Fig. \ref{fig:pol}).
Surprisingly, we find that the experimental data for the distortion
at zero temperature ($u_0$) and above the first magnetic transition
($u_1$) exhibit nearly the same dependence on $r_R$ as either the
orbital polarization $\langle\tau^x\rangle$, or the effective
interaction $g_{\rm eff}$. This is an unexpected outcome of the
present theory as information about the actual lattice distortions has
not been used in constructing the spin-orbital model (\ref{som}), and
implies that:
(i) the full $\vartheta$-dependence of $g_{\rm eff}$ is due to
   $u(\vartheta)$ and the bare coupling parameters $\{g,K\}$ are
   nearly constant and independent of $r_R$ (Fig. \ref{fig:pol});
(ii) $g\simeq 33J$ obtained using $u\simeq 0.030g_{\rm eff}/J$
   (i.e., $g/a_0\simeq 0.15$ eV/\AA{} for $a_0=3.8$ \AA{});
(iii) $\langle\tau^x\rangle=\chi(\vartheta;T) g_{\rm eff}(\vartheta)$,
    where the susceptibility $\chi\simeq 0.2/J$ hardly depends on
    $\vartheta$ and only weakly on $T$ (cf. Fig. \ref{fig:ops}), so that
    $u(\vartheta) \simeq u_0(\vartheta) / [ 1 - \chi(T) g^2 / K ]$,
    which justifies {\it a posteriori\/} our approach with a single
    parameter $g_{\rm eff}$ (\ref{geff}), depending only on $\vartheta$;
(iv) $K>220J$ (as $\chi g^2/K < 1$). $K$ may be estimated from the
shear modulus which is however unknown for $R$VO$_3$. Taking the
data for SrTiO$_3$ \cite{Car07} instead would imply $K\simeq
6600J$ ($\simeq 8$ eV/\AA{}$^2$), i.e., a $3-5$ \% contribution of
$\langle\tau^x\rangle$ to $u$ in $g_{\rm eff}$.

%%%%%%%%%%%%%%%%%%%%%%%%%%%%%%%%%%%%%%%%%%%%%%%%%%%%%%%%%%%%
%%                     open problems
%%%%%%%%%%%%%%%%%%%%%%%%%%%%%%%%%%%%%%%%%%%%%%%%%%%%%%%%%%%%
Finally, we emphasize that the experimental data of
Fig. \ref{fig:phd} are reproduced with rather realistic parameters
--- taking $J=202$ K one finds $T_{N1}=0.73J=147$ K for LaVO$_3$
($T_{N1}^{\rm exp}=143$ K \cite{Miy03}). Although the present
theory brings us closer to the ultimate understanding of the
complex phase diagram of the vanadates, several issues remain
open. One of them is the second phase transition at $T_{N2}$ to
the $G$-AF phase, which occurs for small $r_R$ \cite{Miy03}. As
shown in Ref. \cite{Kha01}, this transition is due to an interplay
between superexchange orbital fluctuations and orbital-lattice
interactions. While this physics is contained in the model
(\ref{som}), its quantitative description including the recent
observations of coexistence of the $G$-AF and $C$-AF order
\cite{Sag06,Sag07} will have to be addressed in future work.

%%%%%%%%%%%%%%%%%%%%%%%%%%%%%%%%%%%%%%%%%%%%%%%%%%%%%%%%%%%%
%%                       conclusion
%%%%%%%%%%%%%%%%%%%%%%%%%%%%%%%%%%%%%%%%%%%%%%%%%%%%%%%%%%%%
Summarizing, we have introduced a microscopic spin-orbital model
that provides a satisfactory description of the systematic trends
for both orbital and magnetic transitions in the $R$VO$_3$
perovskites, including the nonmonotonic behavior of the OO
temperature $T_{\rm OO}$. Thereby the orthorhombic lattice
distortion $u$, which increases from La to Y by one order of
magnitude, plays a crucial role --- it modifies orbital
fluctuations and in this way tunes the onset of both orbital and
spin order in the cubic vanadates.

%%%%%%%%%%%%%%%%%%%%%%%%%%%%%%%%%%%%%%%%%%%%%%%%%%%%%%%%%%%%
%%                    Acknowledgments
%%%%%%%%%%%%%%%%%%%%%%%%%%%%%%%%%%%%%%%%%%%%%%%%%%%%%%%%%%%%
A.M. Ole\'s acknowledges support by the Foundation for Polish Science
(FNP) and by the Polish Ministry of Science and Education Project
No.~N202 068 32/1481.

%%%%%%%%%%%%%%%%%%%%%%%%%%%%%%%%%%%%%%%%%%%%%%%%%%%%%%%%%%%%
%%
%%                       REFERENCES
%%
%%%%%%%%%%%%%%%%%%%%%%%%%%%%%%%%%%%%%%%%%%%%%%%%%%%%%%%%%%%%


\begin{thebibliography}{}

\bibitem{Goo06} J.-S. Zhou and J.B. Goodenough,
                 \prl {\bf 96}, 247202 (2006).

\bibitem{Miy03} S. Miyasaka {\it et al.\/},
                 \prb {\bf 68}, 100406 (2003).

\bibitem{Miy06} S. Miyasaka {\it et al.\/},
                 \prb {\bf 73}, 224436 (2006).

\bibitem{notecg} In the $G$-type phase the order parameter is staggered
               in all directions, while in the $C$-type phase it is
               staggered in $ab$ planes but remains uniform along the
               $c$ axis.

\bibitem{Ulr03} C. Ulrich {\it et al.\/},
                 \prl {\bf 91}, 257202 (2003).

\bibitem{Miy05} S. Miyasaka {\it et al.\/},
                 \prl {\bf 94}, 076405 (2005).

\bibitem{And07} M. De Raychaudhury, E. Pavarini, and O.K. Andersen,
                 \prl {\bf 99}, 126402 (2007).

\bibitem{Kha01} G. Khaliullin {\it et al.\/},
                 \prl {\bf 86}, 3879 (2001).

\bibitem{Ole07} A.M. Ole\'s {\it et al.\/},
                 \prb {\bf 75}, 184434 (2007).

\bibitem{Hor03} P.~Horsch {\it et al.\/},
                 \prl {\bf 91}, 257203 (2003).

\bibitem{Goo07} J.-S. Zhou {\it et al.\/},
                 \prl {\bf 99}, 156401 (2007).

\bibitem{Yan07} J.-Q. Yan {\it et al.\/},
                 \prl {\bf 99}, 197201 (2007).

\bibitem{Goo04} J.-Q. Yan {\it et al.\/},
                 \prl {\bf 93}, 235901 (2004).

\bibitem{Ima04} M. Mochizuki and M. Imada,
                 New J. Phys. {\bf 6}, 154 (2004).

\bibitem{Pav05} E. Pavarini {\it et al.\/},
                 New J. Phys. {\bf 7}, 188 (2005).

\bibitem{Ree06} M. Reehuis {\it et al.\/},
                 \prb {\bf 73}, 094440 (2006).

\bibitem{Goo05} J.-S. Zhou and J.B. Goodenough,
                 \prl {\bf 94}, 065501 (2005).

\bibitem{notevc} The orbital interaction
               ${{\cal K}}_{ij}$ overestimates the
               stability of the $G$-type OO in LaVO$_3$ in the
               cluster approach.

\bibitem{notece} The data for Ce were not included as the mixed-valent
               electronic structure of CeVO$_3$ requires a separate
               study.

\bibitem{Ren03} Y. Ren {\it et al.\/},
                 \prb {\bf 67}, 014107 (2003).

\bibitem{Sag06} M.H. Sage {\it et al.\/},
                 \prl {\bf 96}, 036401 (2006).

\bibitem{Sag07} M.H. Sage {\it et al.\/},
                 \prb {\bf 76}, 195102 (2007).

\bibitem{Fei99} L.F. Feiner and A.M. Ole\'s,
                 \prb {\bf 59}, 3295 (1999).

\bibitem{Silva} T.N. De Silva {\it et al.\/},
                 \prb {\bf 68}, 184402 (2003).

\bibitem{noteos} Orbital correlations along the $c$ axis were renormalized
               to the disordered orbital chain in LaVO$_3$,
               $\langle{\vec\tau}_i\cdot {\vec\tau}_j\rangle=\frac14-\ln 2$.

\bibitem{notetz} The larger/smaller orbital moments
               $|\langle\tau_i^z\rangle|$ alternate along the $c$ axis
               in the $G$-OO phase below $T_{\rm OO}$.

\bibitem{Car07} M.A. Carpenter,
                 American Mineralogist {\bf 92}, 309 (2007).

\end{thebibliography}
\end{document}